\documentclass[showpacs,aps,apl,twocolumn,superscriptaddress]{revtex4-1}

\usepackage{graphicx}
\usepackage{subfigure}
\usepackage{bm}
\usepackage{color}
\usepackage{mathrsfs}

\newcommand{\comment}[1]{\textcolor{blue}{(RE)}}

\begin{document}

\title{Atomistic calculation of the thickness and temperature dependence of exchange coupling through a dilute magnetic oxide}
\author{R. F. L. Evans}
\author{Q. Coopman}
\author{S. Devos}
\author{W. J. Fan}
\affiliation{Department of Physics, The University of York, York, YO10 5DD, UK}
\author{O. Hovorka}
\affiliation{Faculty of Engineering and the Environment, University of Southampton, Southampton SO17 1BJ, UK}
\author{R. W. Chantrell}
\affiliation{Department of Physics, The University of York, York, YO10 5DD, UK}

\begin{abstract}
The exchange coupling of two magnetic layers via a diffuse oxide interlayer is studied with an atomistic spin model. We investigate the effect of magnetic concentration and oxide layer thickness on the effective exchange coupling strength and find an exponential dependence of the coupling strength on the oxide thickness without the need for magnetic pinholes. Furthermore we show that exchange coupling has a strong temperature dependence which is significant for the reversal dynamics during heat assisted magnetic recording.
\end{abstract}
\maketitle

The continuing increase in the data storage capacity of magnetic hard disk drives has led to a comparable decrease in the median grain size in the recording media to satisfy the noise requirements arising from the magnetic recording trilemma\cite{RichterJPDAP2007}. Future reductions in the grain size present a significant challenge due to the thinning of the oxide interlayer separating magnetic grains\cite{Araki, Tamai}. The exchange coupling between grains is an essential design requirement for magnetic recording media due to its effect in stabilizing the long term thermal stability of the recorded information. Recent experiments by Sokalski \textit{et al} \cite{SokalskiAPL2009} measuring the effective exchange coupling showed an exponential thickness dependence of the effective exchange coupling through a thin oxide layer. The interpretation of physical origins of the exchange coupling varies, such as being a result of the presence of magnetic pinholes, magnetostatic ‘orange peel’ effect, the presence of domain walls near the interface, or randomly distributed magnetic atoms within the oxide layer\cite{PlattPRB2000,Pinholes}. The physical origin of the exponential thickness dependence of the exchange coupling is an interesting question. One suggestion is that of magnetic pinholes, where the two magnetic materials are coupled directly, although Sokalski \textit{et al} found no evidence of these in the samples they studied. Another hypothesis is that the exponential dependence arises naturally from a magnetically dilute oxide layer, which contains randomly distributed magnetic atoms due to the sputtering process. In addition to understanding the low temperature behaviour of the exchange coupling, the expected move to heat assisted magnetic recording (HAMR) means that understanding the temperature dependence of the exchange coupling is essential to understanding its effect during the heat assisted write process\cite{HuangKryderJAP2014}.

In this letter we present an atomistic model of the exchange coupling between two ferromagnets coupled through a dilute magnetic oxide which fully supports the scenario of the exponential dependence of the effective exchange coupling resulting from a magnetically diluted oxide layer. We use the model to investigate the effect of interlayer density and thickness on the effective exchange coupling. Finally we find a linear temperature dependence of the effective exchange coupling.

\begin{figure}[!t]\center
\includegraphics[width=7cm, trim=0 0 0 -20]{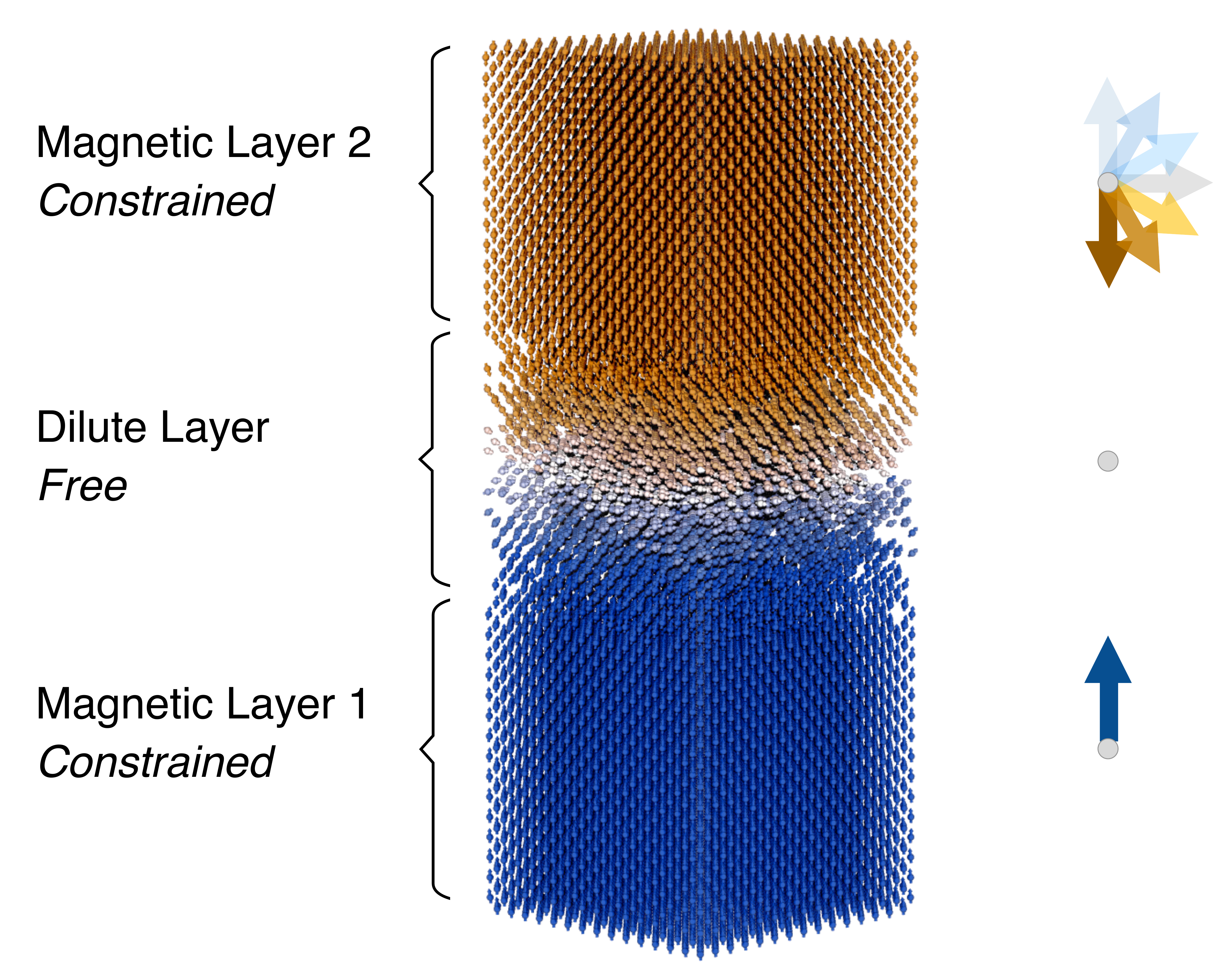}
\caption{Visualization of the simulated system, showing a domain wall between the magnetic layers. Blue/Gold coloring indicates the magnetic orientation along the $\pm z$ axes respectively. For the constrained Monte Carlo simulations, the magnetic layers are held at different angles (indicated by the arrows), while spins in the dilute interlayer are free to point along different directions.(Color Online.)}
\label{fig:vis}
\end{figure}

To model the inter-granular exchange coupling we consider two ferromagnetic layers separated by a dilute magnetic layer with variable thickness and density of atomic spin moments, shown schematically in Fig.~\ref{fig:vis}. The magnetic properties of the systems are simulated with an atomistic spin model\cite{EvansVMPR2013} using the \textsc{vampire} software package\cite{vampire-url}. The energetics of the system of spins is given by the Heisenberg spin Hamiltonian 
\begin{equation} 
\label{eq:Hamiltonian}
\mathscr{H} = -\sum_{i\ne j} J_{ij} \mathbf{S}_i \cdot \mathbf{S}_j 
\end{equation}
where $J_{ij} = 5.6 \times 10^{-21}$ J/link is the exchange interaction between the nearest neighbour sites $i$ and $j$, $\mathbf{S}_i$ is the local spin moment, and $\mathbf{S}_j$ is the spin moment of a nearest neighbouring atom. The spin moments are expressed here as unit vectors $\mathbf{S}_i = \boldsymbol{\mu}_i/|\mu_i|$, where $\mu_i = 1.72 \mu_B$. The parameters chosen for both the magnetic layers and also atoms within the dilute magnetic oxide are representative of Co with a Curie temperature $T_{\mathrm{c}}$ of 1390 K. The simulated system consists of two magnetic layers separated by a dilute magnetic interlayer of variable thickness which controls the coupling between the two layers. The system size is 5 nm $\times$ 5 nm in the $x-y$ plane with periodic boundary conditions in the plane. Each of the fully magnetic layers are 5 nm thick and the dilute interlayer thickness is varied between 1 monolayer and 10 nm. The density of the dilute layer is controlled by randomly removing the desired fraction of atoms from the initial face-centered-cubic crystal, leaving vacancies at some atomic sites. This leads to broken exchange links between atoms in the dilute layer reducing the strength of the exchange coupling. Since we are interested in the effective exchange coupling the spin Hamiltonian omits the usual anisotropy and Zeeman terms as they have no appreciable effect on the calculated exchange coupling energy.

In order to calculate the effective exchange coupling between the two ferromagnetic layers it is necessary to apply a constraint to force and retain a domain wall into the system\cite{AtxitiaEXC2010}. We have developed a hybrid Constrained Monte Carlo/Monte Carlo algorithm (CMC/MC) which allows for the complete control of the domain wall in the system at any arbitrary temperature. The CMC method\cite{AsselinCMC2010} allows the direction of the magnetization of a system to be constrained while fully allowing the length of magnetization to fluctuate thermodynamically. This is achieved by picking random pairs of spins within the constrained set of spins and moving them so that the net direction of the spins is the same but their vector sum is different\cite{AsselinCMC2010}. The hybrid CMC/MC method used for the current work utilizes two constraints, one for each magnetic layer, and a \textit{free} Monte Carlo algorithm\cite{EvansVMPR2013} for the dilute magnetic interlayer, allowing the latter to freely vary its direction and magnitude, as illustrated in Fig.~\ref{fig:vis}.

\begin{figure}[!t]
\includegraphics[width=7cm, trim=25 0 25 0]{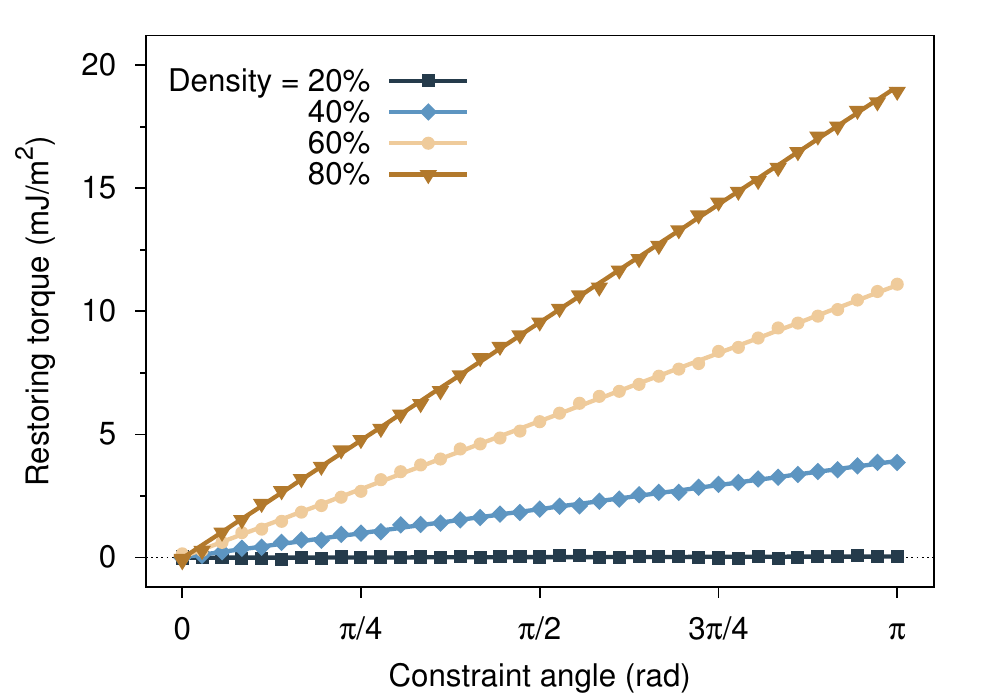}
\caption{Calculated restoring torque as a function of angle between magnetic layers, for different densities of the interlayer for a simulation temperature of $T = 10$ K. The thickness of the interlayer is set at 5 nm. The lines show linear regression fits to the data. (Color Online.)}
\label{fig:TQ}
\end{figure}

To formulate the problem of the calculation of the exchange coupling energy between the two layers, we first consider the energy of the system at the micromagnetic level, averaging over the atomic spins in the system $\mathbf{m} = \langle \mathbf{S} \rangle$. Given two ferromagnetic layers with magnetizations $\mathbf{m}_1$ and $\mathbf{m}_2$, the exchange energy $E_{\mathrm{ex}}$ between the layers can be written as
\begin{equation}\label{eq:Jeff}
E_{\mathrm{ex}} = J_{\mathrm{eff}} \mathbf{m}_1 \cdot \mathbf{m}_2
\end{equation}
where $J_{\mathrm{eff}}$ is the effective coupling between the layers\cite{AtxitiaEXC2010}. Considering parallel and anti-parallel configurations of $\mathbf{m}_1$ and $\mathbf{m}_2$, the energy difference is $2J_{\mathrm{eff}}$. To determine the temperature dependent exchange energy, it is necessary to calculate the $free$ energy difference. Calculation of the \textit{free} energy directly is computationally difficult\cite{AsselinCMC2010}, and so we proceed by calculating the derivative of the free energy (given by the torque) and recovering the \textit{free} energy difference $\Delta F$ by integration\cite{AsselinCMC2010}, such that 
\begin{equation}\label{eq:free}
\Delta F = \int_0^{\theta} \tau \mathrm{d}\theta
\end{equation}
where $\theta$ is the angle between the average magnetizations $\mathbf{m}_1$ and $\mathbf{m}_2$ and $\tau$ is the thermodynamic average of the restoring torque given by
\begin{equation}\label{eq:tq}
\tau = \langle \sum_i \mathbf{S}_i \times -\frac{\partial\mathscr{H}}{\partial \mathbf{S}_i} \rangle \mathrm{.}
\end{equation}
The restoring torque is calculated by fixing the magnetization in the bottom layer along the +$z$ direction, while the top layer is rotated sequentially from the +$z$ axis through 180$^{\circ}$ in steps of 5$^{\circ}$. At each constraint angle the system is first equilibrated for 10,000 Monte Carlo steps (MCS) and then the average restoring torque is calculated for a further 10,000 MCS. The constraint ensures that the domain wall is almost entirely contained within the free layer and so the magnetizations of the constrained layers are virtually uniform. Over the entire system the total torque is zero, since each of the layers exerts an equal and opposite force on the other. The torque is therefore calculated as the difference between the restoring torques on each layer, $\tau = \tau_1 - \tau_2$. The calculated torques for a 5 nm thick interlayer for different interlayer densities are presented in Fig.~\ref{fig:TQ}.

The free energy difference is recovered by integration of the torque, however to improve the accuracy of the calculation, it is clear from Fig.~\ref{fig:TQ} that the calculated torques have a linear dependence in $\theta$. Using a quadrature rule the derivative of the torque with $\theta$ is analytically related to the integral of the torque. The angle dependence of the torque can then be substituted by
\begin{equation}\label{eq:tq-deriv}
\tau = \left(\frac{\mathrm{d}\tau}{\mathrm{d}\theta}\right)\theta 
\end{equation}
where $\mathrm{d}\tau/\mathrm{d}\theta$ is calculated by linear regression of the $\tau(\theta)$ line. The free energy difference between the parallel and anti-parallel configuration of the constrained magnetizations (equal to $J_{\mathrm{eff}}$) is then obtained by combining equations (\ref{eq:Jeff}),  (\ref{eq:free}) and (\ref{eq:tq-deriv}) for a definite integral $\theta = 0 \rightarrow \pi$, giving a final expression for the exchange energy $J_{\mathrm{eff}}$ as
\begin{equation}\label{eq:Jeff-tq}
J_{\mathrm{eff}} = \frac{1}{2}\left(\frac{\mathrm{d}\tau}{\mathrm{d}\theta}\right) \int_0^{\pi}\theta \mathrm{d}\theta = \frac{\pi^2}{4}\left(\frac{\mathrm{d}\tau}{\mathrm{d}\theta}\right)\mathrm{.}
\end{equation}

\begin{figure}[!lt]\center
\includegraphics[width=8cm, trim=0 0 0 0]{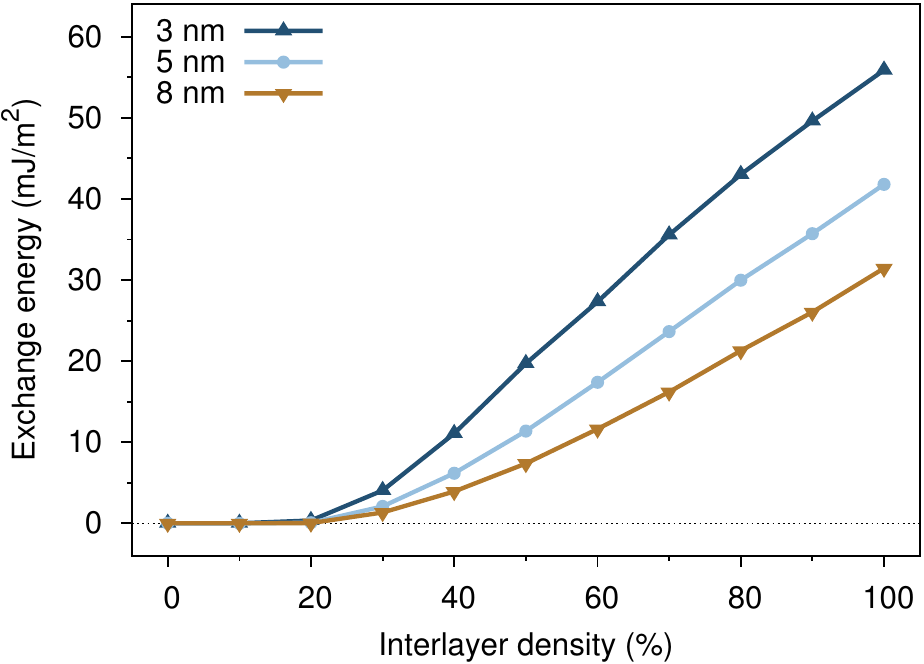}
\caption{Density dependence of the exchange coupling energy for 3 nm, 5 nm and 8 nm interlayer thicknesses at $T = 10$ K. For low densities, the relatively large thickness leads to no coupling between the layers, while at higher densities the coupling is more prominent, approaching the bulk value. As expected increased interlayer thicknesses lead to reduced exchange coupling. (Color Online.)}
\label{fig:density}
\end{figure}

We now proceed to calculate the effective exchange coupling energy, $J_{\mathrm{eff}}$, across the diffuse interlayer. First let us consider the low temperature case and relatively large interlayer thicknesses, as shown in Fig.~\ref{fig:density}. As expected, increased interlayer thickness strongly reduces the coupling at the same density, while increased density beyond a threshold value rapidly increases the coupling energy. The reason for the threshold density for a given thickness lies in the fact that for coupling to exist, there must be a continuous line of atoms between the two layers. For low densities the probability of achieving this is low, but as the density is increased the lines are formed more easily. In a similar fashion, the greater the distance between the layers, the less likely it is that a continuous  line will form, and hence naturally leads to an interlayer thickness dependence of the exchange coupling.

\begin{figure}[!t]\center
\includegraphics[width=8cm, trim=10 0 10 0]{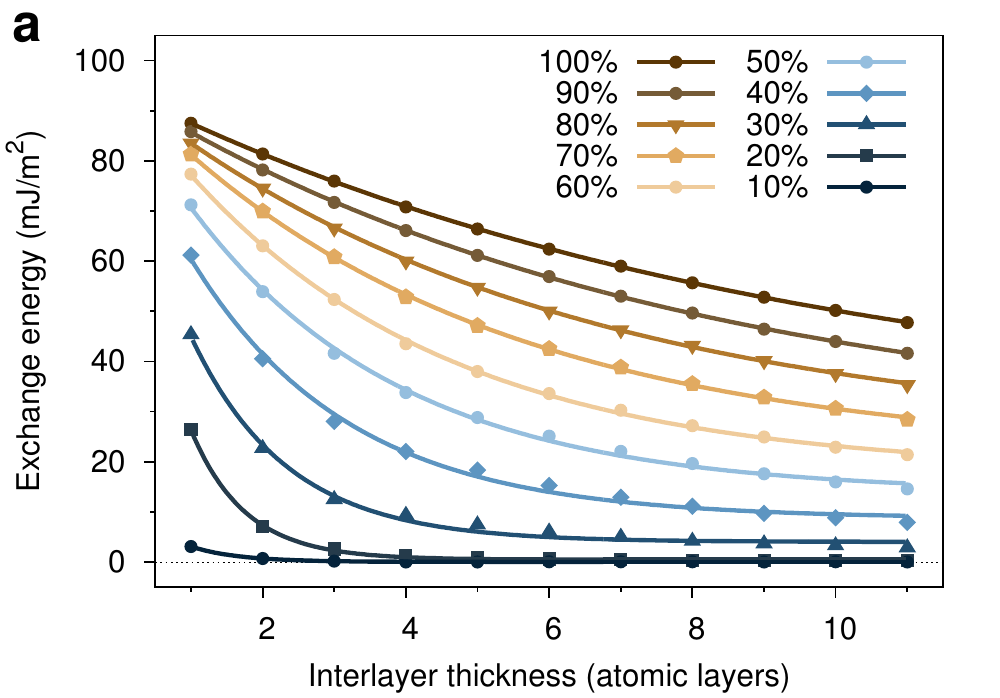}
\includegraphics[width=8cm, trim=10 0 10 0]{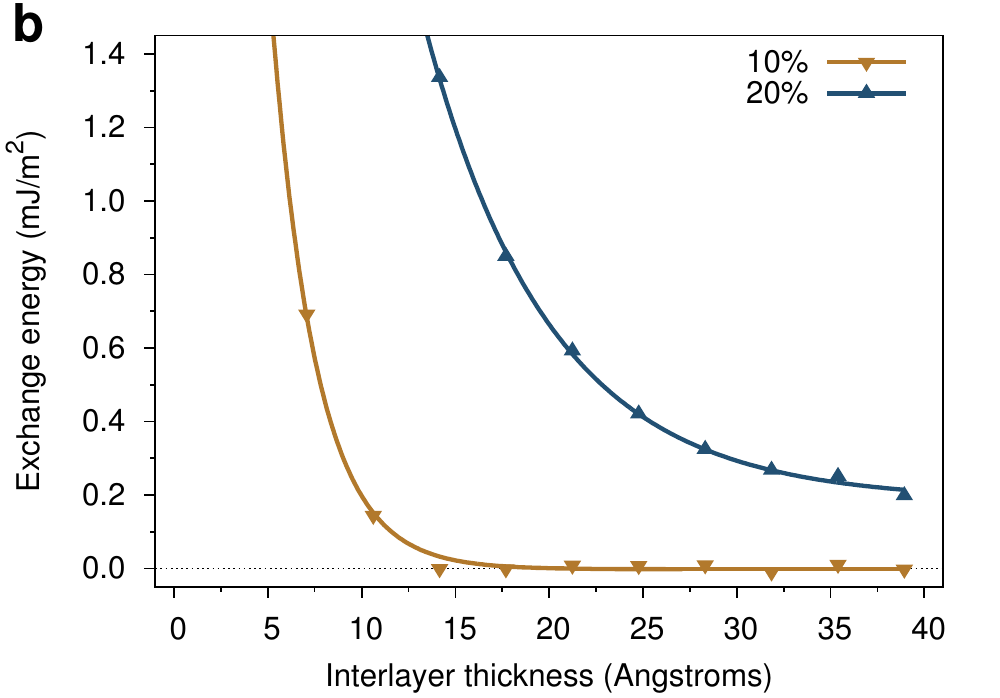}
\caption{(a) Thickness dependence of the exchange coupling energy for different interlayer densities at a temperature of 10 K. (b) Thickness dependence of the exchange coupling energy for low densities comparable with the experimental range from Sokalski \textit{et al}\cite{SokalskiAPL2009}. Lines show exponential fits to the data. For low densities the coupling shows an exponential dependence of the effective exchange coupling, in agreement with experimental results. (Color Online.)}
\label{fig:thickness}
\end{figure}

Although the method works well for a wide range of interlayer thicknesses, the more interesting region is that for thin interlayers, more likely to be found in a granular recording medium. Fig.~\ref{fig:thickness} shows the interlayer coupling strength as a function of the number of monolayers between the magnetic layers for different densities. The exchange coupling follows a non-linear interlayer thickness dependence, which is much stronger for the lower densities. This is again due to the requirement to form a continuous line of atoms between the layers for any effective exchange coupling. Fig.~\ref{fig:thickness}(a) clearly shows that for low densities this coupling is hard to achieve, with almost zero coupling beyond a single monolayer for a density of 10\%. Densities between 20\% and 50\% show a strong exponential thickness dependence as seen experimentally\cite{SokalskiAPL2009}, arising from the increased probability of forming connecting lines of atoms. For densities of 60\% and greater, a different behavior is seen, where the exchange coupling is less critical with interlayer thickness. Here the interlayer thickness dependence of the exchange coupling is no longer dominated by the density of the interlayer, but by the distance over which the domain wall is forced between the two constrained layers. Hence, even 100\% density (continuous) shows a distance dependence, as the angle between adjacent atomic planes is reduced.

For a more direct comparison with the data from Sokalski, Fig.~\ref{fig:thickness}(b) shows details of the thickness dependence of the exchange coupling for low densities, more likely to be seen experimentally. Qualitative agreement with the experimental data is very good, and it can also be seen that the exponential coupling dependence holds, supporting the origin of the coupling as a dilute magnetic material. It is important to note the low temperatures used for the calculations, which gives rise to a non-zero exchange coupling at extended distances, in contrast to the experimental data. At elevated temperatures this extended exchange coupling is reduced due to a low ordering temperature of the interlayer, but suggests that in reality densities of 30\% - 40\% are realistic for the intergranular phase. It is also interesting to note the different exchange coupling found for different Oxide interlayers found by Sokalski. The absorption of metal impurities into the oxide is very likely to be material dependent, and so within the framework of our model different exchange couplings arising from different oxides can easily be explained by a different effective magnetic density of the interlayer due to different degrees of diffusion.

\begin{figure}[!t]\center
\includegraphics[width=8cm, trim=10 0 10 0]{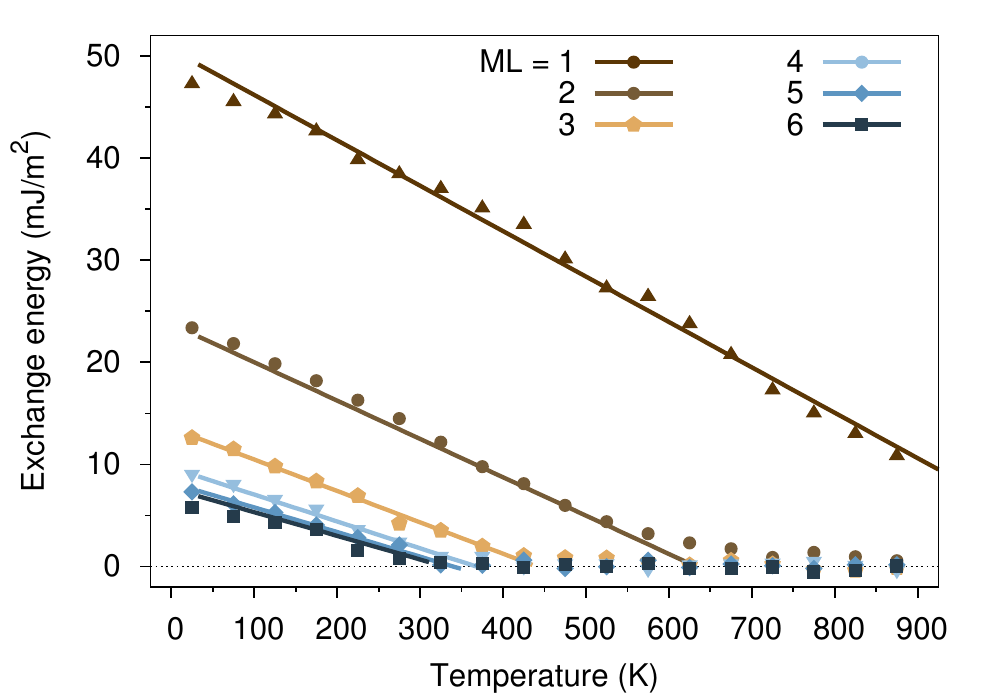}
\caption{Calculated temperature dependence of the exchange coupling energy for different interlayer thicknesses at a density of 30\%. Lines show are from Eq.~\ref{eq:Aeff}. (Color Online.)}
\label{fig:temperature}
\end{figure}

Finally we consider the temperature dependence of the exchange coupling, an effect particularly relevant to heat assisted magnetic recording. In conventional magnetic recording, the temperature dependence above room temperature is of little interest, while for HAMR it is much more significant since the recording medium is heated to a high temperature during the writing process. Fig.~\ref{fig:temperature} shows the simulated temperature dependence of the exchange coupling for different interlayer thicknesses and a constant interlayer density of 30\%. For a single monolayer, the exchange coupling is strong even at elevated temperatures. However, for increasing thicknesses the exchange coupling rapidly decreases due to thermal fluctuations. This is due to the low density of magnetic atoms, which leads to a low intrinsic Curie temperature. For thin layers however, the interlayer is polarized by the ferromagnetic layers, which increases the ordering temperature and stabilizes the exchange coupling. It is also clear that the exchange coupling exhibits a much stronger temperature dependence than the bulk exchange, owing to the dilute nature of the interlayer, and shows a broadly linear temperature dependence. This is in full agreement with recent experimental data for FePt/SiO/FePt trilayers\cite{HuangKryderJAP2014} where a linear dependence of the exchange coupling with temperature was found.
The exchange coupling and effective Curie temperature of the intergranular layer are both exponentially dependent on the interlayer thickness, and so can be conveniently fitted by the function 
\begin{equation}
A_{\mathrm{eff}}(r_{ij}, T) = A(r_{ij})\left(1-\frac{T}{T_{\mathrm{c}}^*(r_{ij})}\right)
\label{eq:Aeff}
\end{equation}
where $A(r_{ij}) = A_0 \exp(-r_{ij}/r_0) + A_{\mathrm{min}}$, $A_0 = 114.08$ mJ/m$^2$ and $A_{\mathrm{min}} = 7.31$ mJ/m$^2$ are fitting constants, $r0 = 1.033$ ML is a characteristic interaction range, $T_c^*(r_{ij}) = T_{\mathrm{c}}^0 \exp(-r_{ij}/r_0) + T_{\mathrm{c}}^{\mathrm{min}}$, $T_{\mathrm{c}}^0 = 2147.14$ K is a fitting constant, and $T_{\mathrm{c}}^{\mathrm{min}} = 323.25$ K is the intrinsic Curie temperature of the interlayer. The fits according to Eq.~\ref{eq:Aeff} are plotted in Fig.~\ref{fig:temperature}. The significant result for HAMR is that the strong temperature dependence of the exchange coupling means that it can be largely ignored during the writing process, and should have a minimal impact on jitter. Thus, the exchange can be engineered towards thermal stability largely without consideration of the hysteric properties of the media.

In conclusion, we have investigated the effect of interlayer thickness, magnetic density, and temperature of the exchange coupling between two magnetic layers. We have shown that pinholes are not necessary to describe an exponential dependence of the exchange coupling and that the coupling is strongly dependent on the density of the interlayer. Furthermore the exchange coupling is strongly temperature dependent suggesting that for HAMR the intergranular exchange is not relevant to the write process, only for long term thermal stability of the media. Given that reduced oxide thickness in future magnetic recording media will likely lead to a greater dispersion of intergranular exchange coupling, alternative approaches for controlling the inter-granular exchange, such as application of a magnetic capping layer\cite{FanJAP2011} may allow better control of the magnetic properties.

This work was supported by the European Community's Seventh Framework 
Programme (FP7/2007-2013) under grant agreement No. 281043 \textsc{femtospin}. and the Advanced Storage Technology Consortium \textsc{astc}. The authors wish to thank Kangkang Wang and Roger Wood for helpful discussions.

\bibliography{library,local}


%
\end{document}